\magnification=\magstep1
\baselineskip=16pt
\def\A{{\cal A}}
\def\S{{\cal S}}
\def\s{ \{|A_k\rangle\} }

\centerline{\bf QUANTUM DISCORD AND MAXWELL'S DEMONS}
\bigskip
\centerline{Wojciech Hubert Zurek}
\centerline{\it Theory Division, T-6, MS B288, LANL, Los Alamos, NM87545}
\bigskip
\centerline{\it Abstract}
\medskip
\noindent {\it Quantum discord was proposed as an information theoretic measure 
of the ``quantumness'' of correlations. I show that discord determines
the difference between the efficiency of quantum and classical Maxwell's
demons in extracting work from collections of correlated quantum systems.}

\bigskip
Information has an energetic value: It can be converted into work. Maxwell's
demon$^1$ was introduced into thermodynamics to explore the role of information
and, more generally, to investigate the place of ``intelligent observers"
in physics. In modern discussions of the subject$^2$ ``intelligence'' is often
regarded as predicated upon or even synonymous with the information processing 
ability -- with computing. Thus, Maxwell's demon is frequently modelled by 
a universal Turing machine -- a classical computer -- endowed with the ability 
to measure and act depending on the outcome. The role of such a demon is then 
to implement an appropriate conditional dynamics -- to react to the state 
of the system as revealed through its correlation with the state of 
the apparatus. It is now known that quantum logic -- i.e., logic
employed by quantum computers -- is in some applications more powerful 
than its classical counterpart. It is therefore intriguing to enquire whether 
a quantum demon -- an entity that can measure non-local states and implement
quantum conditional operations -- could be more powerful than a classical one. 
I show that quantum demons can typically extract more work than classical 
demons from correlations between quantum systems, and that the difference 
is given by a variant of the {\it quantum discord}, recently introduced$^{3-5}$
measure of the ``quantumness'' of correlations.

Quantum discord$^{3-5}$ is the difference between the two classically identical
formulae that measure the information content of a pair of quantum systems.
Several closely related variants can be obtained starting from the original 
definition$^3$ given in terms of the mutual information$^6$. Mutual information 
is a measure of the strength of correlations between, say, the apparatus $\A$
and the system $\S$: 
$$ I(\S:\A) = H(\S) + H(\A)  - H(\S, \A)  \eqno(1)$$
It measures the difference between the missing information about two objects 
when they are taken separately, $H(\S) + H(\A)$, and jointly, $H(\S,\A)$. 
In the extreme case $\S$ and $\A$ are identical -- e.g., two copies of 
the same book, or a state of the apparatus pointer $\A$ after 
a perfect but as yet unread measurement of $\S$. Then the joint entropy 
$H(\S, \A)$ is equal to $H(\A) = H (\S)$, so $I(\S: \A) = H(\A)$. By contrast, 
when the two objects are not correlated, $H(\S,\A) = H(\S) + H(\A)$, and 
$I(\S:\A)=0$. 

The other formula for mutual information employs classical identity for joint entropy$^6$:
$$ H(\S, \A) = H(\A) + H(\S|\A) = H(\S) + H(\A | \S) \eqno(2)$$
Above $H(\S|\A)$ is the conditional entropy -- measure of the lack of knowledge
about the system given the state of the apparatus. Substituting this in Eq. (1) 
leads to an asymmetric looking formula for mutual information:
$$ J(\S:\A) = H(\S) + H(\A) - [ H(\A) + H(\S|\A) ] \eqno(3)$$
We have refrained from carrying out the obvious cancellation above that would 
have yielded $J(\S:\A) = H(\S) - H(\S|\A)$ for a reason that will become 
apparent very soon. 

Discord is defined as:
$$\delta(\S|\A) = I(\S:\A) - J(\S:\A) = 
[ H(\A) + H(\S|\A) ] - H(\S, \A) \eqno(4)$$
Classically, of course, discord disappears as a consequence of Eq. (2). In 
quantum theory the situation is no longer this simple: In order to properly 
define the conditional entropy one must specify how the apparatus is 
``interrogated'' about $\S$: 
Measurements modify the state of the pair. 
After a measurement of the observable with eigenstates $\s$ observer's own 
description of the pair is the conditional density matrix:
$$\rho_{\S\A|A_k\rangle}= \rho_{\S|A_k\rangle} \otimes |A_k\rangle\langle A_k| 
\eqno(5)$$
He will attribute to the system $\rho_{\S|A_k\rangle}$ with the probability 
$p_{\A}(k) = Tr \langle A_k | \rho_{\S\A} | A_k \rangle $.
The post measurement density matrix $\rho_{\S\A}'$ differs form 
the pre-measurement $\rho_{\S\A}$ even for a bystander who has not yet 
found out the outcome. 
This point of view of the bystander differs from the viewpoint of the observer
(or a demon) who made the measurement: Demon knows that the apparatus is in the
state $|A_k\rangle$. Bystander obtains his post-measurement $\rho_{\S\A}'$ by 
averaging over the outcomes. 
$$ \rho_{\S\A} = \sum_k p_{\A}(k) \rho_{\S| A_k\rangle} \otimes
|A_k \rangle \langle A_k| \eqno(6) $$
His description of the pair is unaffected by 
the measurement only when the measured observable commutes with $\rho_{\S\A}$. 
We shall find this bystander viewpoint very useful because it represents 
a statistical ensemble of all possible outcomes.

In quantum physics one possible definition of joint entropy is inspired 
by Eq. (3):
$$ H_{\A}(\S, \A_{\s}) = [H(\A) + H(\S|\A)]_{\s} \eqno(7)$$
where $\s$ is the eigenbasis of the to-be-measured observable of the apparatus.
Another acceptable and completely quantum definition would be to simply 
compute the von Neumann entropy of the density matrix $\rho_{\S\A}$ describing 
the joint state. Then:
$$ H(\S, \A) = - Tr \rho_{\S \A} \lg \rho_{\S, \A}  
= - \sum_l p_{\S\A}(l) \lg p_{\S\A}(l) \eqno(8)$$
where $p_{\S\A}(l)$ are the eigenvalues of $\rho_{\S\A}$ -- the probabilities 
of the density matrix that jointly describes the correlated pair. These 
eigenvalues always exist, but in general correspond to entangled quantum states 
in the joint Hilbert space of $\S$ and $\A$. Such states cannot be found out 
through sequences of local measurements starting with just one subsystem of 
the pair -- say, $\A$. This is a fundamental difference between the quantum 
and the classical realm (where such ``piecewise'' investigation is always 
possible and need not disturb the state of the pair). It is responsible for 
non-zero discord. 

A simple example of this situation is a perfectly entangled state: 
$$ |\psi_{\S\A} \rangle = (|00\rangle + |11\rangle)/\sqrt 2 \eqno(9a) $$
Above, the first entry refers to $\S$ while the second corresponds to $\A$.
Clearly, $\rho_{\S\A} = |\psi_{\S\A} \rangle \langle \psi_{\S\A} |$ is pure 
-- the pair is with certainty in the state $|\psi_{\S\A} \rangle$. Hence, 
$H(\S,\A) = 0$. On the other hand, $\rho_{\A(\S)} = {\bf 1_{\A(\S)}}/2$, 
where ${\bf 1}$ is the unit matrix in the appropriate Hilbert space, so that 
$H(\A) = H(\S) = 1$. Consequently, $I(\S:\A)=2$, but the asymmetric mutual 
information is $J(\S:\A)=1$. This is because the joint information
$H_{\A}(\S,\A_{\s})$ defined 
with reference to any measurement on a $\A$, Eq. (5), is a sum of $H(\A) = 1$ 
and $H(\S|\A)=0$, with both of these quantities independent of the basis 
because of the symmtery of Bell states. 

Readers are invited to verify that a classical correlation in:
$$ \rho_{\S\A} =(|00 \rangle \langle 00|+|11\rangle \langle 11 |)/2 \eqno(9b) $$
results in zero discord, but only when the preferred basis 
${\s} = \{ |0 \rangle, ~ |1 \rangle \} $ is employed. 
The entangled state of Eq. (9a) could be converted into the mixture of Eq. (9b) 
through decoherence in the preferred (pointer) basis$^{4,8-11}$ or -- and this 
is why decoherence can be regarded as monitoring by the environment -- 
through a measurement with an undisclosed outcome carried out in the same 
pointer basis ${\s} = \{ |0 \rangle, ~ |1 \rangle \} $. 

In general, the ignorance of the bystander cannot decrease (but may increase)
as a result of a measurement of a known observable if he does not know 
the outcome$^{12}$. Hence, $H_{\A}(\S, \A_{\s}) - H(\S, \A) \geq 0$, and
$$ \delta(\S| \A_{\s}) \geq 0 \eqno(10)$$
Equality occurrs only when $\rho_{\S\A}$ remains uneffected by a partial 
measurement of $\s$ on the $\A$ end of the pair.

The relevance of the discord for the performance of Maxwell's demon can be now 
appreciated. Demons use the acquired information to extract work from their 
surroundings. The traditional scenario starts with an interaction establishing 
initial correlation between the system and the apparatus. The demon then 
reads off the state of $\A$, and uses so acquired information about $\S$ 
to extract work by letting $\S$ expand throughout the available phase 
(or Hilbert) space of volume (dimension) $d_{\S}$ while in contact with 
the thermal reservoir at temperature $T$. This yields:
$$ W^+ = k_{B_2} T (\lg d_{\S} - H(\S|\A)) \eqno(11) $$
of work obtained at a price:
$$ W^- = k_{B_2} T H(\A) \eqno(12) $$
Above, $k_{B_2}$ is the Boltzmann constant adapted to deal with the entropy 
expressed in bits and $T$ is the temperature of the heat bath. 
The net gain is then:
$$ W = k_{B_2} T (\lg d_{\S} - [H(\A) + H(\S|\A)]) \eqno(13a)$$
The price $W^-$ is the cost of restoring the apparatus to the initial 
ready-to-measure state. The significance of this ``cost of erasure'' for
the second law was pointed out in the seminal paper of Szilard$^{13}$, but its 
relevance in the context of information processing was elucidated and codified 
by Landauer$^{14}$. 

It is now accepted that neither classical$^{15-17}$ nor quantum$^{18-21}$
demons can violate the second law because of the cost of erasure. However,
a demon with a supply of empty memory (used to store measurement 
outcomes) can extract, on the average, $W^+$ of work per step from a thermal 
reservoir. This strategy works, because, in effect, demon is using its memory 
as a reservoir with low entropy. However, a new block of empty memory  of size 
$d_{\A}$ is used up with each new measurement. This is wasteful, and only
fradulent accounting (uncovered by Szilard and Landauer) which ignores 
the thermodynamic value of empty memory can create an appearance of 
the violation of the second law. 

To optimize performance demon should use memory of $\A$ more efficiently. 
The obvious strategy here is to compress the bits of the outcomes after 
a sequence of measurements, freeing up an unused block of length $\Delta \mu$. 
The data can be compressed to the size given by their 
{\it algorithmic complexity}$^{17}$. The savings are:
$$ \Delta \mu = \lg d_{\A} - K(A_k) $$
Where $K(A_k)$ is the algorithmic randomness (Kolmogorov complexity) 
per step. Moreover, one can show that for long sequences of data 
the approximate equality:
$$ \langle K(A_k) \rangle \simeq H(\A) $$
becomes exact, so that the saved up memory is on the average:
$$ \Delta \mu = \lg d_{\A} - H(\A) $$
Maxwell's demon can attain net work gain per step of:
$$ W = k_{B_2} T (\lg d_{\S} d_{\A} - [ H(\A) + H(\S | \A)]) \eqno(13b) $$
When $\S$ and $\A$ are classically correlated so that Eq. (2) applies, this
can be written as:
$$ W = k_{B_2} T (\lg d_{\S} d_{\A} - H(\S , \A)) \eqno(13c) $$
We note that the efficiency is ultimately determined by the joint entropy of 
$\S$ and $\A$ {\it accessible} to the demon, and that the same equation would 
have followed if we simply regarded the $\S\A$ pair as a composite system, and 
the demon used it all up as a fuel. 

The efficiency of demons is then determined by what they know about the pair
$\S\A$ -- its joint entropy -- and we have already seen that in quantum 
physics joint entropy depends on how the information about the pair can be 
acquired. A classical demon is local -- it operates one-system-at-a-time on 
the correlated a quantum pair $\S \A$. In this case the above sketch 
of the ``standard operating procedure" applies with one obvious {\it caveat}: 
It needs to be completed by the specification of the basis demon measures 
in $\A$. The cost of erasure is still given by Eq. (12), also for classical 
demons extracting work from quantum systems$^{11-13}$. Thus: 
$$ W^C/k_{B_2}T = \lg d_{\S \A} -  [H(\A) + H(\S|\A)]_{\s} \eqno(14)$$
The only difference between the classical Eq. (10) and the quantum Eq. (11) is 
the obvious dependence on the basis $\s$ demon selects to measure. 
The expression in the square brackets is the measure of the remaining 
(conditional) ignorance and of the cost of erasure. We shall be interested in 
the ${\s}$ that maximize $W^C$.

A quantum demon can typically extract more work -- get away with lower costs 
of erasure -- because its measurement can be carried out in a basis that avoids
increase of entropy associated with measurements and decoherence$^{4,8-11,23}$: 
It can always select a global basis in the combined Hilbert space of $\S\A$ that
commutes with the initial $\rho_{\S\A}$. The work that can be extracted after 
the apparatus gets reset to its initial ready-to-measure pure state is:
$$ W^Q/k_{B_2} T = \lg d_{\S \A} - H(\S , \A) \eqno(15)$$
The other straightforward way to arrive at Eq. (15) is to use quantum demon in 
its capacity of a universal quantum computer, which, by definition, can 
transform any state in the Hilbert space into any other state. This will, in 
particular, allow the demon to evolve entangled eigenstates of an arbitrary 
known $\rho_{\S \A}$ into product states. The resulting density matrix can be 
measured in a local basis that does not perturb its eigenstates, and, hence, 
as viewed by the bystander, it will not suffer any additional increase of 
entropy. The work extracted by the optimal quantum demon is limited simply 
by the basis-independent joint entropy of the initial $\rho_{\S\A}$.

The difference between the efficiency of the quantum and classical demons
can be now immediately computed:
$$\Delta W/k_{B_2} T =  [H(\A) + H(\S|\A)]_{\s} - H(\S ,\A) \eqno(16)$$
or:
$$\Delta W = k_{B_2} T \delta_{\A}(\S|\A_{\s}) \eqno(17)$$
Equation (17) relating the extra work $\Delta W$ to discord defined as 
the difference of the accessibe joint entropy of classical (local) and quantum 
(global) demons is the principal result of our paper. It answers 
an interesting physics question while simultaneously providing an 
operational interpretation of the discord. 

To gain further insight into implications of the above discussion, let us 
first note that discord is in general obviously basis dependent. Discord 
disappears iff the density matrix has the ``post-decoherence" (or
``post-measurement'') form, Eq. (6), {\it already before the measurement}. 
Given the ability of classical demons to match the quantum performance standard
in this case, basis $\s$ that allowes for the disappearance of discord in the
presence of non-trivial correlation can be justifiably deemed classical. We note
that the $\rho_{\S\A}$ of the locally diagonal form presented above may emerge 
as a consequence of the coupling of $\A$ with the environment$^{8-11}$.
The preferred {\it pointer basis} emerges as a result of einselection.

A typical $\rho_{\S\A}$ does not have the form of Eq. (5), however. In that
case discord does not completely disappear for any basis, and is usually 
basis-dependent. It is therefore of interest to enquire about the basis
that yields the least discord, $\hat \delta(\S|\A)$. This leads us back to 
the ambiguity in the definition of the discord: we could adopt either:
$$ \hat \delta(\S|\A) = min_{\s}[H(\A) + H(\S | \A)]_{\s} - H(\S \A) \eqno(18)$$
or 
$$ \hat \partial(\S|\A) = H(\A) + min_{\s} H(\S | \A)_{\s}-H(\S\A) \eqno(19) $$
The difference between them is obvious, and $\hat \delta(\S|\A)
\geq \partial(\S|\A)$. 

Discord is not symmteric between the two ends of the correlation: In general,
$$\hat \delta(\S\A) \neq \hat \delta(\S\A) \eqno(20)$$ 
In particular, for density matrices that emerge from $\hat \delta(\S|\A)$ may
vanish but $\hat \delta(\A|\S)$ may may remain finite.
Such correlations are {\it one-way classically accessible}. They are 
characterised by a preferred direction -- from $\A$ to $\S$ -- in which
more information about the joint state can be acquired. Thus, a local demon
that can choose between the two ``ends'' of the $\S\A$ pair may be in some
cases more efficient than a one-way demon. Indeed, one could define a
polarization
$$ \varpi(\S|\A) = \hat \delta(\S|\A) - \hat \delta(\A|\S) \eqno(21)$$
to quantify this directionality.

One can generalise discord to situations involving collections of
correlated systems. The obvious strategy is to define it as a difference
between the joint entropy corresponding to a particular sequence of (possibly
conditional) measurements -- that is, the obvious generalisation of Eq. (7)
-- and the joint von Neumann entropy of the unmeasured 
density matrix. One could define a minimal discord of a collection of systems 
as a minimum over all possible sequences of measurements. This corresponds
to the demon having a choice of the end of the pair it can measure first. 

First hints of the quantum underpinnings of the Universe emerged over 
a century ago in a thermodynamic setting involving black body radiation.
We have studied here implications of quantum physics -- and, in particular,
of the quantum aspects of correlations -- for 
classical and quantum Maxwell's demons. We have seen 
that discord is a measure of the advantage afforded by the quantum conditional 
dynamics, and shown that this advantage is eliminated in presence of decoherence
and the ensuing einselection. Our discussion sheds a new light on the problem of
transition between quantum and classical: It leads to an operational measure
of the quantum aspect of correlations. As was already pointed out$^4$,
the aspect of quantumness captured by discord is not the entanglement. Rather, 
it is related to the degree to which quantum superpositions are implicated
in a state of a pair or of a collection of quantum systems. We expect it to be 
relevant in questions involving quantum theory and thermodynamics (see e.g. 
Ref. 24), but discord may be also of use in characterising multiply 
correlated states that find applications in quantum computation. 

\bigskip

\noindent{\bf References}

\item{1.} Maxwell, J. C. {\it Theory of Heat}, 4th ed., pp. 328-329
(Longman's, Green, \& Co., London 1985).

\item{2.} Leff H. S. and Rex, A. F. {\it Maxwell's Demon: Entropy, Information,
Computing}, (Princeton University Press, Princeton, 1990).



\item{3} Zurek, W. H. {\it Annalen der Physik} {\bf 9}, 855 (2000).

\item{4} Zurek, W. H. Decoherence, einselection, and the quantum origin of the 
classical, {\it Rev. Mod. Phys.} (in the press), quant-ph 0105127 (2001)

\item{5} Ollivier, H. and Zurek, W. H. {\it Phys. Rev. Lett.} (2002).

\item{6} Cover, T. M. and Thomas, J. A. {\it Elements of Information Theory}
(Wiley, New York, 1991).

\item{7} Fuchs, C. and Peres, A. {\it Physics Today} {\bf 53}, 70 (2000). 

\item{8} Zurek, W. H., Pointer basis of a quantum apparatus: Into what mixture
does the wavepacket collapse? {\it Phys. Rev.} {\bf D 24}, 1516-1524 (1981). 

\item{9} Zurek, W. H., Decoherence and the transition from quantum to classical
{\it Physics Today} {\bf 44}, 36-46 (1991).

\item{10} Giulini, D., Joos, E., Kiefer, C., Kupsch, J., Stamatescu, I.-O., and 
Zeh, H. D., {\it Decoherence and the Appearance of a Classical World in Quantum 
Theory}, (Springer, Berlin, 1996).

\item{11} Paz, J. P., and Zurek, W. H., Environment-induced decoherence and the
transition from quantum to classical, Les Houches Lectures (2000).

\item{12} Klein, O. {\it Z. Phys.} {\bf 72}, 767-775 (1931).

\item{13} Szilard, L. {\it Z. Phys.} {\bf 53} 840 (1929). English translation in
Behav. Sci. {\bf 9}, 301 (1964), reprinted in {\it Quantum Theory and
Measurement}, edited by J. A. Wheeler and W. H. Zurek (Princeton University
Press, Princeton, 1983); Reprinted in Ref. 2.

\item{14} Landauer, R. {\it IBM J. Res. Dev.} {\bf 3}, 183 (1961).

\item{15} Bennett, C. H. {\it Int. J. Theor. Phys.} {\bf 21}, 905 (1982).

\item{16} Bennett, C. H. {\it Sci. Am.} {\bf 255} (11), 108 (1987).

\item{17} Zurek, W. H. {\it Nature} {\bf 347}, 119-124 (1989);
{\it Phys. Rev.} {\bf A40}, 4731-4751 (1989).

\item{18} Zurek, W. H. Maxwell's demon, Szilard's engine, and quantum 
measurements, pp. 151-161 in {\it Frontiers of Nonequilibrium 
Statistical Physics}, G. T. Moore and M. O. Scully, eds., (Plenum Press, 
New York, 1986); reprinted in Ref. 2.


\item{19} Lloyd, S. Quantum-mechanical Maxwell's demon {\it Phys. Rev.} 
{\bf A56}, 3374-3382 (1997).

\item{20} Nielsen, M. A., Caves, C. M., Schumacher, B., and Barnum, H.
Information theoretic approach to quantum error correction and reversible 
measurement, {\it Proc. Roy. Soc. London} {\bf A454}, 277-304 (1998).

\item{21} Zurek, W. H., Algorithmic randomness, physical entropy, measurements,
and the demon of choice, pp 393-410 in {\it Feynman and Computation}, 
A. J. G. Hey,  ed., (Perseus Books, Reading, 1999).

\item{22} Li, M. and Vitanyi, P. {\it An Introduction to Kolmogorov Complexity
and its Applications} (Springer, Berlin, 1993).

\item{23} Zurek, W. H. Information transfer in quantum measurements, pp. 87-116
in {\it Quantum Optics, Experimental Gravity, and the Measurement Theory},
P. Meystre and M. O. Scully, eds. (Plenum, New York, 1983).

\item{24} Horodecki, R., Horodecki, M., and Horodecki, P. {\it Phys. Rev.}
{\bf A63} 022310 (2001).

\end